\documentclass[sigconf,authorversion]{acmart}
\AtBeginDocument{%
  }

\setcopyright{rightsretained}
\copyrightyear{2025}
\acmYear{2025}
\acmDOI{}
\acmConference[CHI '25 "Everyday AR through AI-in-the-Loop" Workshop]{}{}{}
\acmISBN{978-1-4503-XXXX-X/18/06}

\begin{document}

\title[Do We Need \textit{Responsible XR}?]{
Do We Need \textit{Responsible XR}? Drawing on Responsible AI to Inform Ethical Research and Practice into XRAI / the Metaverse
}

\author{Mark McGill}
\orcid{0000-0002-8333-5687}
\author{Joseph O'Hagan}
\orcid{0000-0003-2905-1584}
\author{Thomas Goodge}
\orcid{0000-0001-6229-7936}
\author{Graham Wilson}
\orcid{0000-0003-2664-1634}
\author{Mohamed Khamis}
\orcid{0000-0001-7051-5200}
\affiliation{%
  \institution{University of Glasgow}
  \city{Glasgow}
  \country{UK}
}

\author{Veronika Krauß}
\orcid{0000-0002-4936-9787}
\affiliation{%
  \institution{University of Applied Sciences Ansbach}
  \city{Ansbach}
  \country{Germany}
}

\author{Jan Gugenheimer}
\orcid{0000-0002-6466-3845}
\affiliation{%
  \institution{TU Darmstadt}
  \city{Darmstadt}
  \country{Germany}
}

\renewcommand{\shortauthors}{McGill et al.}

\begin{abstract}
This position paper for the CHI 2025 workshop "Everyday AR through AI-in-the-Loop" reflects on whether as a field HCI needs to define \textit{Responsible XR} as a parallel to, and in conjunction with, Responsible AI, addressing the unique vulnerabilities posed by mass adoption of wearable AI-enabled AR glasses and XR devices that could enact AI-driven human perceptual augmentation. 
\end{abstract}





\maketitle





\section{XRAI and the Future Augmented Society}
Everyday AR headsets have the potential to supplant our reliance on physical smartphones, monitors and displays \cite{tang-sphere, mcgillExpandingBoundsSeated2020, shady-bans, mmve-joseph}, enabling users to optically and aurally track, understand, and augment the world and its inhabitants. This is likely to herald new capabilities in augmented intelligence \cite{zhengHybridaugmentedIntelligenceCollaboration2017} and perception  \cite{schraffenbergerEverythingAugmentedReal2014, huguesNewAugmentedReality2011,schraffenbergerArguablyAugmentedReality2018}, communication \cite{artanimCreatingInteractiveVR2020, diego-lbw}, productivity \cite{cho2025evaluating, robert-lbw}, accessibility \cite{ar-blind}, and more - promising the transformative ability to ``build a better reality'' \cite{hankeMetaverseDystopianNightmare2021}. AI will be a facilitator and amplifier here, empowering users, communities, business, governments and others to alter, augment, diminish or otherwise mediate our perception of reality \cite{moriSurveyDiminishedReality2017,schraffenbergerArguablyAugmentedReality2018}.




We reflect on the emerging ethical risks \cite{billinghurstGrandChallengesAugmented2021,suggested-wang, slaterEthicsRealismVirtual2020, kraussWellIntentionedHCIStudents2023} and vulnerabilities \cite{bonnailMemoryManipulationsExtended2023,tsengDarkSidePerceptual2022,ruoccoRedirectedNavigationForced2024,kraussWhatMakesXR2023,eghtebasCoSpeculatingDarkScenarios2023} exposed by XR-driven \textit{human perceptual augmentation}, where AR glasses in-particular can instrument our everyday lives, and wearable AI-in-the-loop can then act to alter or override our perception of the world accordingly \cite{ohaganAugmentingPeoplePlaces2023,ohaganViewpointSocietalImpact2024} through personal and metaversal layers atop reality, and ask: do we need \textit{Responsible XR} as a parallel to Responsible AI \cite{tahaeiHumanCenteredResponsibleArtificial2023,mikalefThinkingResponsiblyResponsible2022,arrietaExplainableArtificialIntelligence2020}? And if so, how do we as a community work to define, achieve consensus around, and advocate for best practice given the emerging convergence of consumer wearable spatial computing and AI?

\subsection{The Death of Mental and Bystander Privacy}
A device that can sense, record, and allow AI to ingest the instrumented behaviour, physiology, actions and social interactions of the user inherently undermines mental privacy \cite{yusteItTimeNeuroRights2021,iencaWeHaveRight} through the capacity to model a user’s preferences (biometric psychography), social bias (e.g. revealing aversions or sexual preferences), cognitive and attentional load, mental demand/fatigue and more \cite{ohaganPrivacyEnhancingTechnologyEveryday2023,abrahamDonRecordMy2024,jolie-mum}.  But such devices also risk the privacy of bystanders \cite{ohaganSafetyPowerImbalances2021, reality-aware, chi-joseph, imx-joseph, o2023dynamic}. For example, FRL/MRL posited the concept of \textit{LiveMaps}\footnote{\url{https://www.youtube.com/watch?v=JTa8zn0RNVM}}, arguably a form of distributed public surveillance driven by wearable cameras. 
Privacy is also related to \textit{solitude}, and there is ultimately a question of if we are ever truly alone given wearable, ever-present AI.

\subsection{The Death of a Common, Shared Reality}
AR overlays digital elements onto the physical world, whilst the role of AI is arguably to generate/refine/curate those elements \cite{aghelmaneshHowPeoplePrompt2024,delatorreLLMRRealtimePrompting2024}, tailoring them to individuals \cite{streckerPersonalizedRealityChallenges2024} based on their preference, behaviour and bias, as well as the desires of other stakeholders (platforms, companies, governments etc.) \cite{kraussCreateFearMissing2025,narayanan2020dark}. 
With a single prompt, a user could leverage generative AI to personalize their perception of themselves \cite{bonnerWhenFiltersEscape2023} and their surrounding reality \cite{delatorreLLMRRealtimePrompting2024,aghelmaneshHowPeoplePrompt2024}. This might be for benign or beneficial reasons, such as enhancing mental health by increasing the prevalence of nature on a city walk; or for more questionable purposes, from ``nudification'' \cite{franksDesertUnrealInequality2017} or sexual appropriation of others \cite{ohaganSafetyPowerImbalances2021},  to censoring others based on pre-existing prejudices  \cite{bonnerWhenFiltersEscape2023}.
And more generally, such a technology would, for better or worse, undermine the concept of a shared or even objective reality that we all experience. This poses potential benefits - for example replaying cherished memories tied to a location \cite{bonnailMemoryManipulationsExtended2023} - but also risks  around enacting perceptual filter bubbles or facilitating escapism from reality \cite{eghtebasCoSpeculatingDarkScenarios2023}, and could further transpose the divisions of online life to our everyday perception of reality. Would we be together but apart, perceiving different worlds?


\subsection{The Death of Human Core Skills}
The scope of assistive augmentations to "fundamentally transform human ability" \cite{tanAssistiveAugmentationFundamentally2025} is significant, from enhancing cognition and intelligence \cite{doswellAugmentingHumanCognition2014} to memory \cite{bonnailMemoryManipulationsExtended2023,bonnailWasItReal2024}. However, would we become reliant, even dependent \cite{stewartTechWeRely2024,biswasIncompleteTechEmotional2025}, on such enhancements, and what risk would this pose to core human skills that we develop throughout our lives? A tendency to forget where keys are can be indicative of decreased memory function \cite{singerWhereDidLeave2005} - if our AR glasses remember this for us, does this undermine the utility of such self-report tests of memory, or even degrade our capacity to remember? 

\subsection{The Death of Real Human Communication}
One appealing assistive enhancement is likely to be in AI in the conversational loop - from the functional comprehension of speech (assistive captioning \cite{liAugmentedRealityVisualCaptions2023, mathis2023breaking,mathewAccessDemandRealtime2022}, real-time translation etc.), to better understanding speech content (e.g. providing contextually relevant information) \cite{jadonAugmentedConversationEmbedded2024}, to better understanding the speaker (e.g. their affective state, intonation etc.). Ultimately, this amounts to AI augmenting how we socialise and interact with others. Again, the risk of dependency and a loss of social skills is possible. Consider a first date between two people, both using XRAI conversational assistance - do humans end up the physical means by which two LLMs seduce each other? 

\subsection{When a Place Becomes an Augmented Void}
We are living through the pollution of the internet by AI autophagy in real-time \cite{hallLargeLanguageModels2024,xingWhenAIEats2024}, where AI ingests, generates, distributes and ingests online content once more - an AI Ouroboros arguably leading to homogenization of thought whilst undermining human creativity \cite{georgeErosionCognitiveSkills2024,moonHomogenizingEffectLarge2024,andersonHomogenizationEffectsLarge2024}.
With XR, this could manifest in diminishing physical, tangible creation - consider a public space that becomes an augmented void, where any notable aesthetic or artistic features are purely digital, and where the absence of AR makes the physical environment less desirable \cite{joy2025acceptance}. Indeed, places may become homogenous, subject to the same augmentations, digital displays, and pervasive adverts as illustrated in HYPER-REALITY \cite{keiichimatsudaHYPERREALITY2016}.
%
%

\subsection{Exacerbating Access Inequality}
XRAI demands potentially significant hardware (e.g. \textit{Meta Orion} being projected to cost \$10000) and software (e.g. \textit{ChatGPT} pro currently costs \$200 a month) costs as the economic price to pay for access to such assistive augmentation. This will inevitably lead to further social stratification and a new "digital divide" \cite{tsetsiSmartphoneInternetAccess2017} between the haves and have-nots. Consider interviewing for a job where your competitors have better AI, more seamless perceptual augmentation, and are better practised at operating in synchrony with said technology. Then consider that existing socioeconomic disparities would ensure that select nations would benefit from such human augmentation preferentially over emerging nations.

\subsection{The Death of Agency and Autonomy}

XR also innately exposes users to new vulnerabilities around deception and manipulation \cite{kraussWhatMakesXR2023,ruoccoRedirectedNavigationForced2024}, particularly if coerced in an attempt to avoid the aforementioned access inequality through subsidised access to headsets, metaversal platforms, and the AI that drives XR experiences \cite{kraussWhatMakesXR2023,ohaganAugmentingPeoplePlaces2023}. 
%
Given the technology's capacity to instrument its user and the world around them, and mediate their perception of said world, the value to be extracted from subsidised access would likely lie around data - erosion of mental privacy and worldscraping in particular - and control - over our perception \cite{ohaganAugmentingPeoplePlaces2023}, attention \cite{pandjaitanAuctentionARAuctioningVisual2024}, cognition and memory \cite{bonnailMemoryManipulationsExtended2023} and our resultant thoughts, behaviours, attitudes and actions. Ultimately, those stakeholders that subsidise or gatekeep access to these powerful technologies might choose to diminish our autonomy and agency over the decisions we take in life to better serve their aims - from directing purchasing \cite{ruoccoRedirectedNavigationForced2024} to influencing political views \cite{skwarekAugmentedRealityActivism2018}. XRAI would then mirror what we have seen occur in social media - the commodification of thought and behaviour enacted through the algorithms that determine what we perceive in our \textit{onlife} \cite{szakolczaiOnlifeCriminology}.
%
Kasahara \textit{et al.} \cite{kasahara2024generative} presented a closed-loop system with the intention to unconsciously influence a user’s cognitive processes or even decision-making. They propose a combination of a generative adversarial system combined with reinforcement learning that connects the user in a closed-loop system (using fMRI) to an generative image generator. The images are generated with the intention of inducing a specific mental state in the user and gradient ascent on the latent space is used to steer these adaptations in the right direction. They discuss their system as a traditional desktop application, but if combined with everyday XR and the images generated are filters on top of reality, which influences the user imperceptibly, the result becomes a “perfect” manipulation machine presenting a very powerful and dangerous concept: \textit{Computational Perceptual Manipulations} in XR.

\subsection{When the Virtual Becomes Too Real}
Finally, there is consideration of the risks of ever-increasing perceptual realism \cite{slaterEthicsRealismVirtual2020,madaryRealVirtualityCode2016}, interactional realism \cite{wilsonViolentVideoGames2018} and plausibility \cite{slaterPlaceIllusionPlausibility2009} fueled by generative AI \cite{chamolaRealityPivotalRole2024}, particularly in the near-term for immersive VR, but of increasing performance as we reach high-fidelity everyday AR. From moral risks of isolation and withdrawal from reality \cite{ramirezRealMoralProblems2018}, to the potential for real affective impact and trauma based on perceptually realistic unreal experiences or simulations \cite{podoletzCriticalReviewVirtual2024}, there is a consideration as to whether it remains ethical to pursue genuine perceptual realism.

\section{The Need to Define \textit{Responsible XR}?}

The vignettes above cherry pick potential emergent harms envisioned around the mass adoption of everyday XRAI, but are illustrative that the balance between benefit and harm for this emerging, intersectional technology is at stake. 
We posit that HCI in particular needs to play a more active, strategic, and communal role in considering these risks and the technological, social and policy mitigations that could safeguard society from the worst of this.
%
%
%
%
%
We need to take responsibility for the ethical and responsible exploration and dissemination of research around emerging technology harms. Is it responsible to use scenario elicitation or design fictions  \cite{mhaidliIdentifyingManipulativeAdvertising2021,kraussWhatMakesXR2023,eghtebasCoSpeculatingDarkScenarios2023,ruoccoRedirectedNavigationForced2024} to map out future vulnerabilities and harms, or is this providing a roadmap to those that might exploit this technology? Should we as a community be considering responsible use/ethics statements \cite{kraussWhatMakesXR2023} or is this paying lipservice to ethics? 
Moreover, how as a community can we more effectively come together to define and understand the risks posed \cite{slaterSpeculationEthicsVirtual2021}, similar to related efforts such as the \textit{AI Risk Repository} \cite{slatteryAIRiskRepository2024}? How can we foster interdisciplinary consideration (e.g. Criminology, Law) of these risks? 
And how can we distil an ethical and responsible vision for how XR could work (e.g. considering reliability, safety, trust) that could see the backing of industry and government? One route might be to consider \textit{perceptual rights} \cite{ohaganAugmentingPeoplePlaces2023,ohaganViewpointSocietalImpact2024} that might enshrine limitations to how perceptual mediation could be exploited, similar to how e.g. the EU AI Act begins to limit risky AI and AI's capacity for manipulation.

\begin{acks}
This research was supported by UK Research and Innovation (UKRI) under the UK Government’s Horizon Europe funding guarantee (\href{https://augsoc-project.org/}{AUGSOC}) [EP/Z000068/1]. It was also supported by the AI Safety Institute (AISI) under the \textit{WearAI} project.
\end{acks}

\bibliographystyle{ACM-Reference-Format}
\bibliography{main}

\end{document}